\documentclass[sigplan,10pt]{acmart}
\renewcommand\footnotetextcopyrightpermission[1]{}
\usepackage{graphicx}
\usepackage{balance}  

\usepackage{cleveref}
\usepackage[inline]{enumitem}

\usepackage{url}

\usepackage[utf8]{inputenc}
\usepackage{booktabs}
\usepackage{subcaption}

\usepackage{multicol}
\usepackage{multirow}
\usepackage{ctable}
\usepackage{listings}
\usepackage{xcolor}
\usepackage{hyperref}
\hypersetup{
    colorlinks=true,
    linkcolor=black,
    filecolor=black,      
    urlcolor=black,
    citecolor=black,
}
\usepackage{float}

\AtBeginDocument{%
  \providecommand\BibTeX{{%
    \normalfont B\kern-0.5em{\scshape i\kern-0.25em b}\kern-0.8em\TeX}}}

\begin{document}

\title{Micro-architectural Analysis of a Learned Index}

\author{Mikkel Møller Andersen}
\email{mikka@itu.dk}
\affiliation{%
  \institution{IT University of Copenhagen}
  \country{Denmark}
}
\author{Pınar Tözün}
\email{pito@itu.dk}
\affiliation{%
  \institution{IT University of Copenhagen}
  \country{Denmark}
}

\begin{abstract}

Since the publication of \textit{The Case for Learned Index Structures} in 2018 \cite{originalLearnedIndex},
there has been a rise in research that focuses on learned indexes for different domains and with different functionalities.
While the effectiveness of learned indexes as an alternative to traditional index structures
such as B+Trees have already been demonstrated by several studies,
previous work tend to focus on higher-level performance metrics such as throughput and index size.
In this paper,
our goal is to dig deeper and investigate how learned indexes behave at a micro-architectural level compared to traditional indexes. 

More specifically, we focus on previously proposed learned index structure ALEX \cite{alex},
which is a tree-based in-memory index structure that consists of a hierarchy of machine learned models.
Unlike the original proposal for learned indexes, ALEX is designed from the ground up to allow updates and inserts.
Therefore, it enables more dynamic workloads using learned indexes.
In this work, we perform a micro-architectural analysis of ALEX and
compare its behavior to the tree-based index structures that are not based on learned models,
i.e., ART and B+Tree.

Our results show that ALEX is bound by memory stalls,
mainly stalls due to data misses from the last-level cache.
Compared to ART and B+Tree,
ALEX exhibits fewer stalls and a lower cycles-per-instruction value across different workloads.
On the other hand,
the amount of instructions required to handle out-of-bound inserts in ALEX
can increase the instructions needed per request significantly (10X) for write-heavy workloads.
However, the micro-architectural behavior shows that 
this increase in the instruction footprint exhibit high instruction-level parallelism,
and, therefore, does not negatively impact the overall execution time.

\end{abstract}

\settopmatter{printfolios=true}
\settopmatter{printacmref=false}
\maketitle



\section{Introduction}
\label{intro}

Learned index structures have emerged as an alternative to traditional database indexes in the recent years
since their potential was demonstrated by Kraska et al. \cite{originalLearnedIndex}.
The fundamental idea behind learned indexes is that one can treat the index as a model that predicts the position of a given key in the dataset.
A key advantage of this type of index comes as a result of the reduced index size,
which can even be orders of magnitude reduction in memory footprint,
since the trained model or the hierarchy of models in the index keep less space than keeping actual data. 
In addition, such indexes trade-off increased computations to reduced memory accesses,
which could boost performance if the overall computation cycles are cheaper with respect to memory access latency.
On the other hand, for such an index to be effective in its predictions,
the index model has to be trained on the target dataset prior to deployment,
which limits the dynamic nature of the index. 

There has been several proposals \cite{histtree, alex, pgmindex, fittingtree, shifttable, radixspline, learnedMultiDim}
for learned index structures since the proposal from Kraska et al. \cite{originalLearnedIndex}.
These proposals and initial benchmarking efforts \cite{benchmarkingTUMMIT, benchmarkingSOSD}
already compare learned indexes to index structures that aren't based on learned models. 
However, these comparisons are done by mainly focusing on higher-level metrics such as
throughput (requests completed per second), latency (average time to complete a request), index size, scalability with respect to number of threads utilized, etc.
While these metrics are extremely important and necessary to understand the overall benefits and disadvantages of learned index structures,
they are one side of a coin, especially for main-memory-optimized index structures.
These metrics by themselves do not give a detailed understanding of how learned indexes utilize the micro-architectural resources of a processor
(e.g., utilization of the front-end resources and the different levels of the cache hierarchy).
In this paper, our goal is to shed light on this other side of the coin. 

To achieve our goal, we focus on the learned index ALEX \cite{alex},
which is one of the first learned indexes that is designed ground up to enable efficient updates and inserts.
We perform a micro-architectural analysis of ALEX,
while comparing its behavior to two index structures that are not based on learned models;
ART \cite{artpaper, artcode}, a trie-based index structure, and
B+tree \cite{btreecode}, a tree-based index structure.
The reason we pick these two structures to compare against specifically is that
ALEX is already compared against them in the original paper \cite{alex} and by other studies \cite{benchmarkingTUMMIT, histtree}.
This would allow our results to be cross-checked.

We create a benchmark driver
\footnote{Will be made available once the paper is accepted.}
that generates YCSB-like \cite{ycsbpaper} workloads.
Using this driver, we generate four workloads with varying read-write intensities and two datasets of different sizes. 
We then report the breakdown of execution cycles into different micro-architectural components following
Intel's Top-down Micro-architecture Analysis Method (TMAM) \cite{tmam}.
Our study demonstrates the following:

\begin{list}{\labelitemi}{\leftmargin=1.5em}
    \item{Similar to ART and B+Tree, ALEX spends majority of its execution cycles in memory-bound stalls as a result of the long-latency data stalls from the last-level cache. In contrast to ART and B+Tree, in general, ALEX exhibits fewer stalls and spends fewer cycles per instruction across different workload patterns.}
    \item{On the other hand, inserts to the right-end of the tree (\textit{out-of-bound} keys) lead to an order of magnitude increase in the per-request instruction footprint of ALEX, while the impact for ART and B+Tree is not significant, due to model re-training. The instruction-level parallelism for such instructions, however, is quite high. Therefore, this increase in instruction footprint does not significantly impact the overall execution time for ALEX.}
\end{list}

The rest of the paper is organized as follows.
First, \Cref{background} surveys related work and
gives a brief summary of the index structures used in this study as well as the micro-architecture of a processor.
Then, \Cref{setup} presents the experimental setup and methodology, and \Cref{results} presents and discusses the results of the analysis.
Finally, \Cref{conclusion} concludes with a summary of our findings and sketches directions for future research.

\section{Background and Related Work}
\label{background}

Before diving into our experimental study,
\Cref{background:indexes} presents a brief background on the index structures we use in this paper,
\Cref{background:uarch} describes the micro-architectural components of a modern commodity out-of-order (OoO) processor to aide in the discussion of the results, and
\Cref{background:related} surveys related work on benchmarking learned indexes and micro-architectural analysis of data-intensive systems.

\subsection{ALEX, ART, B+Tree}
\label{background:indexes}

\textbf{ALEX} \cite{alex} is an in-memory updatable learned index based on the recursive model index (RMI) introduced by Kraska et al. \cite{originalLearnedIndex}.
There is a hierarchy of index nodes similar to B+Tree, but each node keep a linear regression model instead of keys to direct the tree search. 
During a key lookup, from root to leaves, the model at each level predicts the child node to go to in the lower level based on the search key.
Once a leaf (data) node is reached, the model at this node predicts the position of the key being searched. 
If the prediction was off, exponential search is used to find the actual key position.
During a key insert, the same traversal steps are used to first predict where the key should be inserted.
If the prediction is not correct, then an exponential search determines the correct insert position.
The core idea is that if models are mostly correct,
meaning their predictions are either correct or very close to the correct position,
an exponential search would be faster than the binary search routine.
In addition,
ALEX uses gapped arrays, leaving space in between data items, to help with inserts
and supports structural modification operations to adapt the tree dynamically. 
In this paper, we use the open-source implementation of ALEX \cite{alexcode},
which is provided by the authors of the original ALEX paper \cite{alex}.

\textbf{ART} (Adaptive Radix Tree) \cite{artpaper} is a space-efficient general-purpose trie-based index structure designed for in-memory database systems.
To reduce the height of the tree and also save space, ART adopts techniques like a high fanout (max 256),
lazy expansion, path compression, and different node types for more compact node layouts.
This leads to better cache efficiency and improves overall performance. 
In this paper, we use the open-source implementation of ART \cite{artcode},
which is provided by Armon Dadgar,
who does not have any affiliation with the authors of the original paper \cite{artpaper}.
This open-source codebase is also used by Ding et al. \cite{alex} while comparing ALEX to ART.

\textbf{B+Tree} \cite{graefeBtree} is the traditional index structure that is at the core of most database management systems.
Over the years, there has been several proposals for different variants of B+Trees
to optimize them for modern processors, in-memory- or SSD-optimized systems, etc. 
The B+Tree implementation used in this paper is provided by Timo Bingmann \cite{btreecode},
which is the same codebase used by Ding et al. \cite{alex} while comparing ALEX to B+Trees
as well as by related work that compares learned indexes to B+Trees \cite{histtree, benchmarkingTUMMIT}.

\subsection{Micro-architecture of an OoO processor}
\label{background:uarch}

The cores of an out-of-order processor has two building blocks at a high-level:
\textit{\textbf{front-end}} and \textit{\textbf{back-end}}.

\textit{\textbf{Front-end}} consists of micro-architectural components
that are responsible from processing \textit{instructions}.
More specifically, it handles fetching, decoding, and issuing instructions.
Fetching instructions require going through the memory hierarchy,
which is composed of three levels of caches (L1, L2, L3) and main memory (DRAM) on most commodity server hardware.
\textit{L1 instruction cache (L1-I)}, which is private for each core,
is responsible from keeping instructions closer to the core.
Ideally, the instructions to be executed next should be found in L1-I.
If they miss in L1-I,
then they have to be fetched from lower-levels of the memory hierarchy,
which has higher latency and stalls the entire instruction pipeline.
In practice, however,
an OoO processor is equipped with various mechanisms to prevent or overlap such stalls.
For example, the \textit{decoder} component,
which decodes fetched instructions into micro-operations ($\mu$Ops),
can process multiple instructions in a cycle.
It is extremely important to prevent front-end from stalling,
since it would naturally lead to stalling or underutilization of the back-end as well.

\textit{\textbf{Back-end}} consists of micro-architectural components
that are responsible from the execution of $\mu$Ops issued by the front-end.
These $\mu$Ops are registered to the \textit{reservation station},
which tracks the operands and the dependencies for $\mu$Ops
and forwards them to the corresponding \textit{execution unit}.
Execution units can operate in parallel unless there are dependencies across,
which enables further instruction-level parallelism. 
Execution units also have private buffers to buffer outstanding data load/store requests,
which allows overlapping stall time that may happen due to these requests.
For example, if a data load request misses from \textit{L1 data cache},
same as the case for instructions,
one needs to fetch the data from lower-levels of the memory hierarchy,
which incurs higher latency and may stall the back-end if not overlapped.
When all the operands are in place for the $\mu$Op,
it is finally executed and \textit{retired}.

Overall,
modern processors adopt several methods to provide implicit parallelism
and overlap the stall time due to different operations
to prevent underutilization of both the front-end and back-end components.
In addition,
\textit{instruction} and \textit{data prefetchers}, \textit{branch prediction}, and \textit{speculative execution}
aim at fetching and processing the instructions and data to be needed
before they are needed to avoid stalling the instruction execution pipeline as much as possible. 
Despite all these techniques, for complex data-intensive systems,
it is well-known that majority of the execution time may be spent in stalls
due to instructions or data cache misses as the next section covers. 

\subsection{Related work}
\label{background:related}

Workload characterization of data-intensive workloads
using micro-architectural analysis is a widely-used method to understand
both how well commodity processors serve data-intensive applications and
how well data-intensive systems utilize commodity processors. 
Prior work on micro-architectural analysis range from investigating the behavior of
online transaction processing (OLTP) \cite{keeton98, SirinTPA16, stets02, AtoE},
to online analytical processing (OLAP) \cite{whereDoesTimeGo, hardavellas07, partha98, SirinOLAP},
to larger-scale cloud \cite{sparkUarch, clearingTheClouds, warehouseUarch, cloudsuiteUarch} workloads and systems.
Overall conclusion from these studies is that many widely-used data-intensive systems underutilize modern commodity hardware,
barely reaching an instructions-per-cycle value of one where the theoretical maximum is four,
even though the main cause of this underutilization may change from system to system and workload to workload.

Based on the findings of these studies, 
computer architects can improve modern server hardware to target the needs of popular data-intensive applications,
and the designers of data-intensive systems can adopt techniques or re-design systems to make their software more hardware-conscious. 
For example,
it was well-known that traditional OLTP systems suffered from front-end stalls due to L1 instruction cache misses
since they had long and complex instruction footprints. 
Sirin et al. \cite{SirinTPAA21} show that the improved instruction fetch unit of Intel's Broadwell architecture
reduces the instruction related stall time for OLTP workloads compared to running them on Intel's IvyBridge.
In addition,
the leaner system design of in-memory OLTP systems leads to a smaller instruction footprint per transaction
and minimize the impact of the stall time spent in fetching instructions.

All the studies mentioned above look at data management systems as a whole
and do not investigate the behavior of their index structures in isolation.
This work, in contrast, performs a finer-granularity analysis focusing solely on the index structures themselves.
At this finer-granularity,
Kowalski et al. \cite{admsTreeProfiling} performed a micro-architectural analysis of two modern OLTP indexes
(BwTree \cite{levandoski2013the} and ART \cite{artpaper}) focusing on the impact of locality.

Our work is orthogonal to all these work as we investigate how well a learned index structure utilize
the micro-architectural resources of a modern processor in comparison to non-model based in-memory indexes.
As learned indexes are fundamentally different than the index structures that are not based on machine learning models,
the findings of the previous studies are not representative for learned indexes. 
As learned indexes are fairly new, this aspect of their performance has not been studied in detail yet.
Our work is an additional step in understanding the impact and characteristics of learned indexes.

The most comprehensive benchmarking work for learned indexes is done by Marcus et al. \cite{benchmarkingTUMMIT}.
This work compares three flavors of learned indexes to tree-based, trie-based, hash-based, etc. index structures.
The authors also touch upon the impact of caching and cache misses at a high-level.
However, they do not perform a detailed breakdown of stalls into different micro-architectural components or cache levels.
Therefore, our work is complementary to this study.

\section{Experimental Methodology and Setup}
\label{setup}

As we highlighted in the previous sections,
our high-level goal in this paper is to complement related work and
observe how learned indexes utilize the micro-architectural components of a modern commodity processor.
To achieve this goal, we ask the following questions:

\begin{list}{\labelitemi}{\leftmargin=1.5em}
    \item{Where does time go when we use learned indexes? Are execution cycles mostly wasted on memory stalls or used to retire instructions?}
    \item{Are memory stalls mainly due to instruction or data accesses?}
    \item{How much instruction-level parallelism can learned indexes exploit?}
    \item{How much do the data size, the workload type, and data access and insertion patterns impact the answers to the questions above?}
    \item{How different is the behavior of the learned indexes in comparison to older non-model based index structures?}
\end{list}

The following subsections describe our experimental methodology and setup to answer these questions.

\subsection{Systems used}
\label{setup:systems}

\subsubsection{Software}
We scope our experimental study to in-memory-optimized indexes for OLTP workloads.
Based on our methodology, this work could be easily expanded to indexes optimized for different workloads. 
As mentioned in \Cref{intro} and detailed in \Cref{background:indexes},
we specifically picked learned index ALEX \cite{alexcode}, ART \cite{artcode}, and B+Tree \cite{btreecode}.
ALEX \cite{alex} is a state-of-the-art proposal for an adaptive updatable learned index.
Therefore, it fits very well for our scope.
Comparing it to ART and B+Tree helps with making our results comparable to existing studies \cite{alex, benchmarkingTUMMIT, histtree}.
The open-source codebases used for ALEX and B+Tree are written in cpp and ART is written in c.

\subsubsection{Hardware}
All the experiments were run through Intel DevCloud \cite{devcloud}
on an Intel Xeon processor (from the family of Intel's Scalable Processors),
which is representative of commodity server hardware used for many data-intensive applications and systems.
The full parameters of the server can be seen in \Cref{specs}. 
The OS installed on this server was Ubuntu 18.04.4 LTS.

\begin{table}
    \centering
    \begin{tabular}{|c|c|}
        \hline
        \shortstack{Processor \\ \phantom{Empty}} & \shortstack{Intel(R) Xeon(R) \\Platinum 8256 CPU}\\\hline
        Clock Speed & 3.80GHz\\\hline
        No. of sockets & 2\\\hline
        Issue width & 4\\\hline
        Cores per socket & 4\\\hline
        L1 data cache & 32 KB\\\hline
        L1 instruction cache & 32 KB\\\hline
        L2 cache & 1 MB \\\hline
        L3 cache (shared) & 16.5 MB\\\hline
        Main memory & 192 GB\\\hline
    \end{tabular}
    \caption{Specifications of the server.}
    \label{specs}
\end{table}

\subsection{The benchmark driver}
\label{setup:driver}

To compare the three index structures, 
we have a lightweight benchmark driver written in cpp for each index.
This driver creates a YCSB-like \cite{ycsbpaper} workload mixes to mimic key-value system requests considering the index entries as key-value pairs.
More specifically, there are four types of requests:
(1) \textbf{\textit{read}}ing a key-value pair's value given a key,
(2) \textbf{\textit{update}}ing key-value pair's value given a key,
(3) \textbf{\textit{insert}}ing a new key-value pair, and
(4) \textbf{\textit{delete}}ing a key-value pair.
Implementation-wise, we directly use the read, insert, etc. interfaces provided by the index codebases.

Each experiment run using this driver is composed of four phases:
(1) the \textbf{\textit{data population}} with the given input characteristics and size,
(2) the \textbf{\textit{warm-up}} with the given number of requests,
(3) the \textbf{\textit{workload run}} with the given mix of read, update, insert, delete requests, and
(4) the \textbf{\textit{wrap-up}} with the result summary statements.
All the reported results in \Cref{results} is measured from the third phase of the experiments.

The key arguments given as input to the benchmark driver to generate customized workloads are the following:
(1) the number of initial key-value pairs to populate the index with;
(2) the number of requests to run after the initial data population and warm-up phases are over;
(3) the distribution of requests among reads, inserts, updates, and deletions;
(4) the upper and lower bound for the input keys in read requests during the workload run, where the keys to be looked up are picked at random; 
(5) the upper and lower bound for keys to be inserted; 
(6) and the flag to indicate whether to create keys to be inserted in consecutive order or randomly both during population phase and workload run.
Values can be created based on a range as well, but the exact value for them is less crucial.
Finally, the data type for keys and values are currently hand-coded to 64bit unsigned integer,
and the random access pattern is uniform across the range given.

\begin{table}
\centering
\begin{tabular}{|l|c|c|c|}
\hline

\textbf{Pattern} & Consecutive & Consecutive & Random \\ \hline \hline

\multicolumn{4}{|l|}{\textbf{Keys after data population}} \\ \hline
\textbf{\# keys} & 160 million & 1.6 billion & 1.6 billion \\ \hline
\textbf{lower range} & 0 & 0 & 0 \\ \hline
\textbf{upper range} & 160 million & 1.6 billion & 3.2 billion \\ \hline \hline

\multicolumn{4}{|l|}{\textbf{Workload run}} \\ \hline
\textbf{\# requests} & 160 million & 1.6 billion & 1.6 billion \\ \hline

\end{tabular}
\caption{Parameters related to data and workload generation for the three sets of experiments.}
\label{bench}
\end{table}

\subsection{The generated workloads}
\label{setup:workload}

\Cref{bench} gives a summary of the options we used to generate the workloads for the experiments in \Cref{results}.
Overall, we generate three sets of experiments.
These sets of experiments help us observing the impact of 
both data creation and access patterns and data sizes.

In the first two sets, we have a \textbf{\textit{consecutive}} data pattern, where we experiment with two dataset sizes.
As also mentioned above, this means that keys are both populated and inserted in consecutive order. 
This case is good for having a dense balanced index structure right after data population.
In addition,
for a learned index like ALEX,
this helps in generating precise models right after index population,
which benefits read-heavy scenarios.
However, it could stress the indexes for the insert-heavy cases,
since the inserts are all out of bounds of the current learned range.
The \textit{consecutive} case also helps us with only requesting keys that exist in the index in read requests,
which is performed at random based on the already known key range.

To contrast this, we have the \textbf{\textit{random}} key generation both for data population and inserts during the workload run based on a given range,
which would help ALEX with the inserts because of its adoption of gapped arrays (see \Cref{background:indexes}).
In this scenario, the read requests during the workload run may ask for keys that do not exist in the index.
We picked the upper key range in this scenario large enough to generate a good random pattern for inserts,
but small enough to avoid too many reads for keys that do no exist in the index.

Within these three sets, we generate four workload types by changing the distribution of different request types:
(1) \textbf{\textit{Read only}} consisting of 100\% reads;
(2) \textbf{\textit{Read heavy}} consisting of 80\% reads, 10\% updates, and 10\% inserts;
(3) \textbf{\textit{Write heavy}} consisting of 40\% reads, 30\% updates, 20\% inserts, and 10\% deletes; and
(4) \textbf{\textit{Write only}} consisting of 100\% inserts.


The warm-up phase for each experiment is 100,000 random read requests.
We report results from a single run of each experiment in \Cref{results},
where total number of requests are in millions or billions as reported by \Cref{bench}.
Because of our large sample size during runs, we did not repeat the experiments.
On the other hand, during our test runs with smaller samples,
we did not observe any pathological cases or outliers.
Finally, all the experiments are single-threaded.
While this is partially the limitation of the codebases we are using,
this is not a limitation for the main goal of this study.
When it comes to investigating how well a codebase utilizes the micro-architectural resources of a core,
focusing on the best-case scenario, where the program can utilize a single core without waiting for other concurrent threads, is desired. 

\subsection{Reported metrics and how they are measured}
\label{setup:metrics}

To align ourselves with Intel's Top-Down Micro-architecture Analysis Method (TMAM) \cite{SirinTMAM, tmam},
we organize our metrics into four levels going from the top-level metrics to finer-granularity lower-level metrics:
\textit{overall performance},
\textit{breakdown of execution cycles},
\textit{breakdown of back-end stalls},
\textit{breakdown of memory stalls}.
Next, we detail these levels as well as TMAM.
The summary of all metrics can be found in \Cref{metrics}.

\subsubsection{Overall performance}

These metrics quantify the behavior of indexes we focus on at a high-level.

\textbf{\textit{Memory footprint}}
reports the total memory used by the indexes after an experiment with a particular workload is over.
This is measured with Linux \textit{top} command \cite{top},
which was set to report results every second and write the output to a file.
We selected the last updated value before the experiment ended to report in this paper.
This reported value contains the memory footprint of both the benchmark driver and the indexes.
However the benchmark driver only stores a few variables.
Therefore, its memory footprint is negligible compared to the index.
We also confirmed this by looking at the statistics reported by ALEX codebase, which outputs the size of the index,
and comparing what was reported by ALEX to what we got from \textit{top}.
This metric gives us an idea about the data footprint of the different index structures.

\textbf{\textit{Execution time}}
reports the average time it takes to execute a workload request (e.g., lookup, insert).
Our benchmark driver uses \texttt{std::chrono::high\_resolution\_clock} to 
capture the time it takes to complete the total number of requests issued 
by a workload after the indexes are already populated with the initial dataset and a warm-up period has passed.
Then, this total execution time is divided to total number of requests issued to get the average execution time per request.
We confirmed the negligible impact of our benchmark driver to the overall execution time
by checking the percentage of time it takes using Intel VTune Profiler's Hotspot analysis \cite{vtune}.
This metric gives us an idea about the overall efficiency of the different index structures.

\textbf{\textit{Instructions retired per request}}
reports the average number of instructions retired to execute a workload request (e.g., lookup, insert).
The total number of instructions retired from running a particular experiment is collected from VTune.
Vtune allows us to instrument the benchmark driver code to collect this information only for the workload run phase of the experiments
using Instrumentation and Tracing Technology APIs \cite{ittapis}.
We, then, divide this total number with the total number of requests issued in the experiment.
We are aware that the benchmark driver itself contributes to the total number of instructions retired,
but the impact is negligible as mentioned under \textit{Execution Time}.
This metric gives us an idea about the instruction footprint of the different index structures.

\textbf{\textit{Cycles per instruction}}, which is the inverse of instructions retired per cycle,
reports the average number of execution cycles it takes to retire instructions during a workload run.
It is calculated by dividing the total cycles used to total number instructions retired, which are both reported by VTune.
It gives us an idea about how much instruction-level parallelism the different index structures exhibit.
The lower this value is the higher instruction-level parallelism a program has, spending on average fewer cycles per instruction.
The theoretical minimum for this value on the hardware used in our experiments is 0.25,
since the modern Intel processors can issue (and retire) up to four instructions in a cycle.

\begin{table}
    \centering
        \begin{tabular}{|l|l|}
        \hline
        \multicolumn{2}{|l|}{Level 1: Overall Performance}       \\ \specialrule{.2em}{.1em}{.1em}
        \multirow{4}{*}{} & Memory footprint     \\ \cline{2-2} 
                          & Execution time   \\ \cline{2-2} 
                          & Instructions retired per request   \\ \cline{2-2} 
                          & Cycles per Instruction  \\ \hline \hline
        \multicolumn{2}{|l|}{Level 2: Breakdown of execution cycles}        \\ \specialrule{.2em}{.1em}{.1em}
        \multirow{4}{*}{} & Retiring    \\ \cline{2-2} 
                          & Bad speculation  \\ \cline{2-2} 
                          & Front-end bound   \\ \cline{2-2} 
                          & Back-end bound   \\ \hline \hline
        \multicolumn{2}{|l|}{Level 3: Breakdown of back-end stalls}     \\ \specialrule{.2em}{.1em}{.1em}
        \multirow{2}{*}{} & Core bound  \\ \cline{2-2} 
                          & Memory bound  \\ \hline \hline
        \multicolumn{2}{|l|}{Level 4: Breakdown of memory stalls}     \\ \specialrule{.2em}{.1em}{.1em}
        \multirow{5}{*}{} & L1 bound    \\ \cline{2-2} 
                          & L2 bound  \\ \cline{2-2} 
                          & L3 bound   \\ \cline{2-2} 
                          & DRAM bound   \\ \cline{2-2}
                          & Store bound   \\ \hline
        \end{tabular}
    \caption{Top-down analysis metrics (with levels listed from top to bottom).}
    \label{metrics}
\end{table}

\subsubsection{Breakdown of execution cycles}

It is extremely difficult to perform a precise breakdown of execution cycles
into where they come from at the micro-architectural level on a modern OoO processor. 
As \Cref{background:uarch} mentions,
modern processors adopt various techniques to enable implicit parallelism within a core
and overlap the stall cycles due to various micro-architectural components.
Earlier studies \cite{whereDoesTimeGo, clearingTheClouds, SirinTPA16, AtoE}
took the approach of using hardware counters to measure cache misses
from different levels of the cache hierarchy and then multiplied them 
with the associated latency for a  miss from that particular level.
While this method gives an idea of instruction or data access bottlenecks,
it ignores the overlaps and may attribute a higher cost to a particular miss than in reality.
Therefore, in 2014,
folks from Intel proposed \textit{Top-Down Micro-architecture Analysis Method (TMAM)} \cite{vtuneTMAM, tmam},
where the new-generation Intel processors were expanded with hardware event counters 
and a methodology for using them to more precisely determine
which micro-architectural components cause instruction pipeline to stall. 
In this methodology, the micro-architectural components are viewed top-down,
and the execution time is broken down into these components
starting from front-end and back-end (\Cref{background:uarch}) at the highest-level and then diving deeper.
Sirin et al. \cite{SirinTMAM} report precisely which counters are used to calculate the different levels and components of TMAM,
and Intel's VTune Profiler \cite{vtune} provides an API to perform this analysis, which we also utilize in this work.

Following TMAM, we also break down execution cycles into four components at this level:
\textit{Retiring}, \textit{Bad speculation}, \textit{Front-end bound}, and \textit{Back-end bound}.

For a CPU with an issue width of four, like the one used in our experiments,
the cores have four pipeline slots meaning that for each clock cycle the front-end can fill the pipeline slots with up to four $\mu$Ops
and the back-end can retire up to four $\mu$Ops.
When there are no stalls,
$\mu$Ops can either retire, contributing to the \textbf{\textit{Retiring}} category,
or they do not retire; mainly due to branch misprediction, which is attributed to the \textbf{\textit{Bad speculation}} category.

If a pipeline slot is empty during a cycle it will be counted as a stall.
The stall will be attributed to the front-end, \textbf{\textit{front-end bound}},
if the front-end was unable to fill the pipeline slot; for example due to instruction cache misses.

If the front-end is ready to deliver the $\mu$Ops, but the back-end is not ready to handle it then the stall is attributed to the back-end,
\textbf{\textit{back-end bound}}.
In the case that both the front-end is not able to deliver a $\mu$Op and the back-end is not ready to handle it, then the stall will be attributed to the back-end.
The reason is that if the the front-end was optimized and therefore ready to deliver the $\mu$Ops,
there would still be a stall since the back-end is not ready to handle it.

\subsubsection{Breakdown of back-end stalls}

The stalls due to front-end and back-end can be further broken into their components.
In this work, we report only breakdown of the back-end stalls based on two reasons.
First, the \textit{back-end bound} is the major component of stalls for the index structures we evaluate (see \Cref{results}).
Second, the \textit{front-end bound} typically hints large instruction footprint and poor code layout;
it can be attributed to the instruction related misses for the most part.

The back-end stalls can be of two main components: \textit{Core bound} or \textit{Memory bound}.
\textbf{\textit{Core bound}} stalls represent the stalls due to data/resource dependencies,
whereas \textbf{\textit{memory bound}} stalls are mainly due to data misses.

\subsubsection{Breakdown of memory stalls}

Lastly one can identify which level of the memory hierarchy the \textit{memory bound} stalls originate from,
by digging deeper into the \textit{memory bound} category from the breakdown of the back-end stalls.
\textbf{\textit{L1 bound}} category represents stalls despite hitting the L1 cache,
which could be due to data TLB misses, pressuring the L1 data cache bandwidth due to a large number of requests, etc.
\textbf{\textit{L2 bound}}, \textbf{\textit{L3 bound}}, and \textbf{\textit{DRAM bound}} categories, respectively,
represent stalls due to data misses that miss in L1, L2, and L3, but hit in L2, L3, and DRAM.
Finally, \textbf{\textit{Store bound}} category represents
how often the core encountered a stall on store operations because the store buffer is full.

At this level, the measurements are less precise than the higher-levels,
meaning that slight overlaps in stall time could be double counted \cite{pipVSticks}.
This leads to the situation that sum of \% breakdown values reported by VTune
will not necessarily match with their parent (one higher-level) value.
However, they do still give a good indication of where the bottlenecks are. 
In this paper,
we normalized these metrics to match the parent-level \textit{Memory bound} \% with the formula below,
where $M$ is the metric we want to normalize, e.g. L1 bound. 
\[ M_{norm} = \frac{M\cdot(memory \ bound)}{L1+L2+L3+DRAM+Store} \]




\section{Experimental Results}
\label{results}

Following the setup and methodology described in the previous section,
we now present the results of our experiments to answer the questions listed at the beginning of \Cref{setup}.
We group the results in four parts following the four levels of metrics \Cref{setup:metrics} describes.
First, \Cref{results:overall} presents the results for overall high-level performance metrics.
Then, \Cref{results:bkExec}, \Cref{results:bkBE}, and \Cref{results:bkMB}
perform a top-down micro-architectural analysis and
break down the execution cycles into increasingly finer-granularity components.

\subsection{Overall Performance}
\label{results:overall}

\subsubsection{Memory footprint}
\label{results:overall:memoryfootprint}

We first focus on the memory footprint of the indexes,
which \Cref{size} reports.
As \Cref{setup:metrics} explains,
the reported numbers are from the end of the workload runs
including the impact of the newly inserted keys during those runs on the index size.
Therefore, the footprint, and the size, of the index for the \textit{read-only} workload is the smallest for all cases. 

From \Cref{size},
we see that ALEX overall has slightly larger memory footprint compared to B+Tree, 
except for the case of \textit{insert-only} workload.
On the other hand, both B+Tree and ALEX are approximately half the size of ART.
This trend is actually different than what is reported by Ding et al. \cite{alex}, especially for B+Trees.
However, we are looking at the whole memory footprint of indexes,
rather than just the index size, which may impact our conclusions.
We also have slightly different workloads and data patterns. 

When we insert \textit{out-of-bound} keys,
meaning that the new key is higher than the current largest key value,
the memory footprint of all indexes grow drastically.
Keep in mind that our experiments are based on total number of requests and not time as \Cref{setup:workload} describes.
Therefore, we know that indexes have exactly the same amount of keys inserted at the end of the \textit{insert-only}
experiments for the \textit{consecutive} pattern.
The memory footprint of ALEX is twice of its initial size after the \textit{insert-only} workload.
B+Tree gets an even bigger hit with tripling in size.
ART's footprint also more than doubles for the initial dataset size of 160 million keys,
but when the initial size is already big (1.6 billion keys), the growth is smaller (approx. 50\%).

Finally, the scenario with \textit{random} inserts behave slightly differently.
ALEX does not exhibit a growth except for the \textit{insert-only} case,
because in the random scenario the inserts could be handled easily by the gapped array structure of ALEX (\Cref{background:indexes}),
and do not trigger heavy structure modification operations.
Overall, all the indexes exhibit a smaller growth in their memory footprint for the \textit{insert-only} scenario.
We suspect this is due to fewer structural modification operations triggered in general with the \textit{random} pattern.

\begin{table}
    \centering
        \begin{tabular}{|l|l|l|l||l|}
        \hline
        \multicolumn{2}{|l}{}      &     \multicolumn{2}{|l||}{Consecutive} & Random \\ \hline
        \multicolumn{2}{|l|}{ALEX}      & 160M keys & 1.6B keys & 1.6B keys \\ \specialrule{.2em}{.1em}{.1em}
        \multirow{4}{*}{} & Read Only   & 3.706  & 34.431 & 34.448 \\ \cline{2-5} 
                          & Read Heavy  & 4.158  & 38.482 & 34.448 \\ \cline{2-5} 
                          & Write Heavy & 4.564  & 42.534 & 34.448 \\ \cline{2-5}
                          & Insert Only & 7.805  & 74.765 & 46.109  \\ \hline
        \multicolumn{2}{|l|}{ART}       & 160M keys & 1.6B keys & 1.6B keys \\ \specialrule{.2em}{.1em}{.1em}
        \multirow{4}{*}{} & Read Only   & 8.128  & 80.417 & 70.506 \\ \cline{2-5} 
                          & Read Heavy  & 8.628  & 85.185 & 73.328 \\ \cline{2-5} 
                          & Write Heavy & 8.652  & 85.415 & 74.137 \\ \cline{2-5}
                          & Insert Only & 20.270  & 125 & 93.267 \\ \hline
        \multicolumn{2}{|l|}{B+Tree}    & 160M keys & 1.6B keys & 1.6B keys \\ \specialrule{.2em}{.1em}{.1em}
        \multirow{4}{*}{} & Read Only   & 3.100  & 28.603 & 28.603 \\ \cline{2-5} 
                          & Read Heavy  & 3.731  & 34.638 & 46.379 \\ \cline{2-5} 
                          & Write Heavy & 4.363  & 40.673 & 49.519 \\ \cline{2-5}
                          & Insert Only & 9.426  & 88.953 & 61.927 \\ \hline
        \end{tabular}
    \caption{Memory footprint for the different indexes at the end of each experiment in GB.}
    \label{size}
\end{table}

\begin{figure*}
\centering
\begin{subfigure}{0.3\linewidth}
\centering
\includegraphics[width=\linewidth]{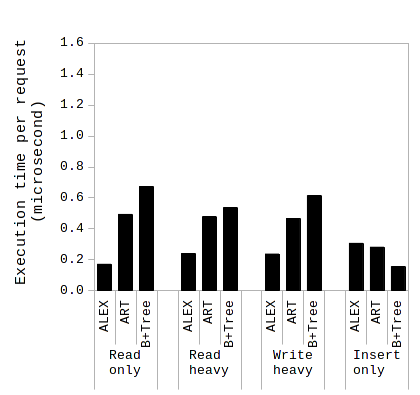}
\caption{Consecutive - 160 million keys}
\label{ExecutionTime_160M}
\end{subfigure}%
\hspace{1em} 
\begin{subfigure}{0.3\linewidth}
\centering
\includegraphics[width=\linewidth]{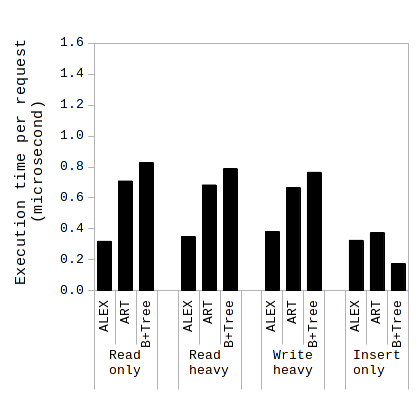}
\caption{Consecutive - 1.6 billion keys}
\label{ExecutionTime_1.6B}
\end{subfigure}
\hspace{1em} 
\begin{subfigure}{0.3\linewidth}
\centering
\includegraphics[width=\linewidth]{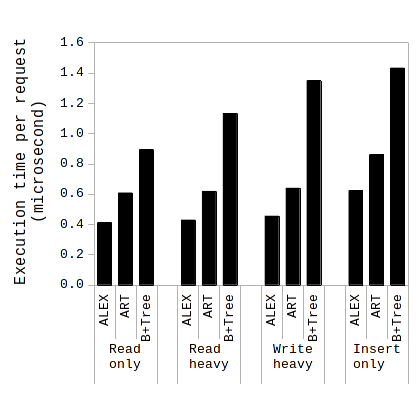}
\caption{Random - 1.6 billion keys}
\label{ExecutionTimeR_1.6B}
\end{subfigure}
\caption{Execution time per request in $\mu$sec.}
\label{ExecutionTime}
\end{figure*}

\subsubsection{Execution Time}
\label{results:overall:executiontime}

\Cref{ExecutionTime} plots the \textit{execution time} per request in $\mu$secs for all the experiments. 
\Cref{ExecutionTime_160M} and \Cref{ExecutionTime_1.6B} focus on the experiments,
where the keys are consecutive from the given range as \Cref{setup:workload} detailed.
Except for the \textit{insert-only} workload,
ALEX outperforms both ART and B+Tree in all workload cases,
completing requests at least in half the time it takes for ART and B+Tree.
This is as expected and corroborates the results from Ding et al. \cite{alex}.
The consecutive keys help with the model predictions,
especially in read-only and read-heavy cases for ALEX.

On the other hand,
inserting keys consecutively from out-of-bound values,
is a costly operation for ALEX as the index has to expand the root.
Normally, when ALEX expands a node, it either scales or retrains its model
and then re-insert all the elements into the new expanded node
based on the prediction of the scaled/retrained model.
As described by Ding et al. \cite{alex}, though,
after some point, if ALEX realizes the append-only behavior in inserts,
it stops model-based re-insertions, and just appends to the expanded space.
The advantage of this is to avoid, re-modeling and re-inserting frequently.
This routine leads to interesting micro-architectural behavior, as the following subsections will cover.

Overall,
while the average execution time per request decreases
as we increase the insert in ART and B+Tree in the consecutive case,
we do not observe this for ALEX,
whose relative performance becomes similar to or worse than ART and B+Tree.
In addition, the increase in data size,
cause an increase in the average execution time for requests for each index,
which is expected as more data is traversed. 

In the case of \textit{random} routine, \Cref{ExecutionTimeR_1.6B},
where the initial key range and later inserts are all randomly done from a given range,
we observe a different trend.
ALEX always outperforms the other two indexes and shows fairly stable performance across different workloads,
except for a slight increase in the average execution time for the insert-only workload.
Compared to the consecutive dataset, the overall execution time is slightly higher for ALEX with random keys.
We see similar trends for B+Tree,
but its average execution time per request exhibits a performance hit with random keys.
For ART, on the other hand, except for the random insert-only workload,
the average execution time per request is lower in the random case if we compare results for 1.6 billion keys
from \Cref{ExecutionTime_1.6B} and \Cref{ExecutionTimeR_1.6B}. 

In the \textit{random} scenario,
the workloads with read requests have a non-negligible possibility of requesting a key that does not exist in the index
(see \Cref{setup:workload}).
Therefore, this pattern leads to higher execution time per request in ALEX in two ways.
First, the models could be less precise due to more random distribution of keys.
Second, we pay the penalty of increased number of exponential search routines for non-existing keys
compared to random read request in \Cref{ExecutionTime_1.6B}.
In addition, in the case of ART, 
we know that a successful search in radix trees is slower than an unsuccessful one
as mentioned by Leis et al. \cite{artpaper}.
This explains ART's opposite behavior compared to ALEX and B+Tree in this scenario.

\begin{figure*}
\centering
\begin{subfigure}{.3\linewidth}
\centering
\includegraphics[width=\linewidth]{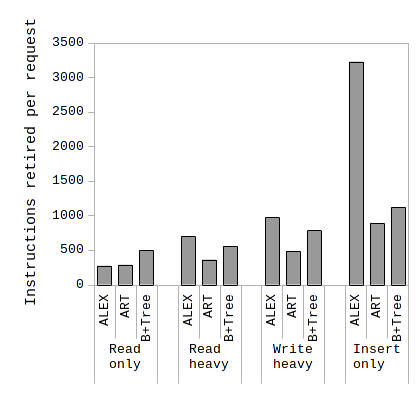}
\caption{Consecutive - 160 million keys}
\label{Instructions_160M}
\end{subfigure}%
\hspace{1em} 
\begin{subfigure}{.3\linewidth}
\centering
\includegraphics[width=1\linewidth]{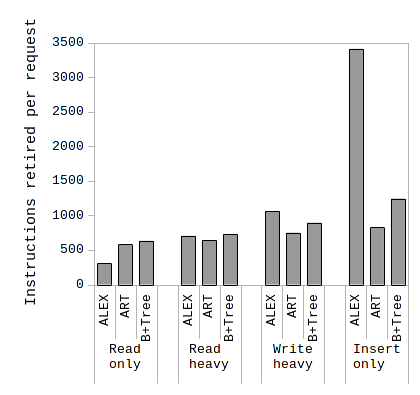}
\caption{Consecutive - 1.6 billion keys}
\label{Instructions_1.6B}
\end{subfigure}
\hspace{1em} 
\begin{subfigure}{.3\linewidth}
\centering
\includegraphics[width=1\linewidth]{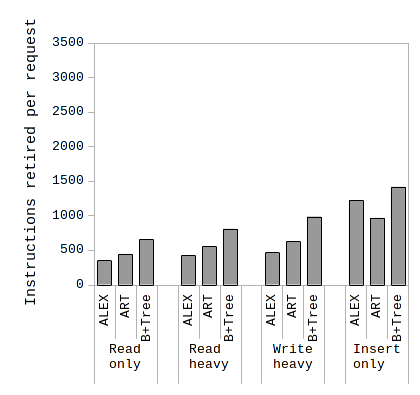}
\caption{Random - 1.6 billion keys}
\label{InstructionsR_1.6B}
\end{subfigure}
\caption{Instructions retired per request.}
\label{Instructions}
\end{figure*} 

\subsubsection{Instructions retired per request}
\label{results:overall:instr}

\begin{figure*}
\centering
\begin{subfigure}{0.3\linewidth}
\centering
\includegraphics[width=\linewidth]{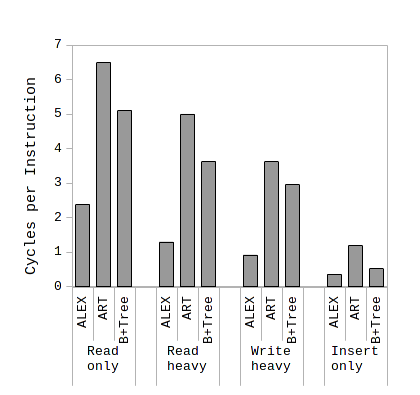}
\caption{Consecutive - 160 million keys}
\label{CPI_160M}
\end{subfigure}%
\hspace{1em} 
\begin{subfigure}{0.3\linewidth}
\centering
\includegraphics[width=\linewidth]{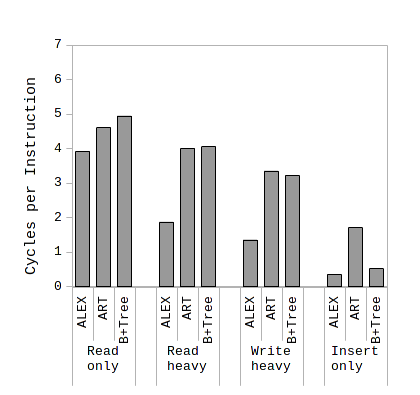}
\caption{Consecutive - 1.6 billion keys}
\label{CPI_1.6B}
\end{subfigure}
\hspace{1em} 
\begin{subfigure}{0.3\linewidth}
\centering
\includegraphics[width=\linewidth]{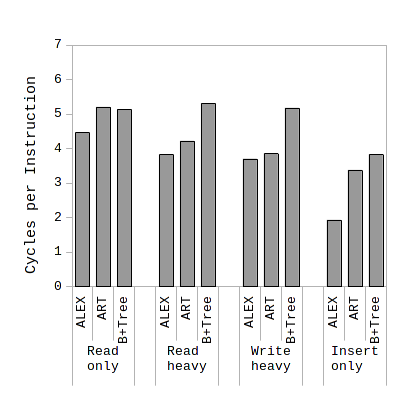}
\caption{Random - 1.6 billion keys}
\label{CPIR_1.6B}
\end{subfigure}
\caption{Cycles per Instruction - CPI (theoretical minimum is 0.25 on our server).}
\label{CPI}
\end{figure*}

\Cref{Instructions} has the instructions retired per request across all experiments
to give an idea of the instruction footprint of the three indexes.
What jumps out from \Cref{Instructions_160M} and \Cref{Instructions_1.6B} is that on average
ALEX has an order of magnitude increase in the number of instructions it uses
to complete an out-of-bound insert request compared to a read operation.
This is due to retraining of the model after so many inserts in the workload.
Even if ALEX stops re-modeling and re-inserting
after detecting the append-only pattern as mentioned in \Cref{results:overall:executiontime},
it may have already performed a model update before that stop happens.
As a result, this impacts the average execution time
as well as average instruction footprint of such insert operations.
On the other hand, as we will see in \Cref{CPI} in the next section,
these instructions exhibit a high instruction-level parallelism,
which, in turn, prevents a drastic increase in the overall execution time as we see in \Cref{ExecutionTime}.

Overall,
the increase in the \% of inserts in the workload increase the instruction footprint for all indexes,
but the impact is not as drastic as it is in the case of ALEX.
In addition, for the case of \textit{random} inserts,
presented in \Cref{InstructionsR_1.6B},
ALEX also gets effected less
as the overheads of such inserts can be prevented by the gapped array structure.

When comparing \Cref{Instructions_160M} and \Cref{Instructions_1.6B},
we also observe that increasing the initial index size 10-fold naturally has impact on instruction footprint per request, 
as it may take longer time to traverse the tree.
This impact is not very high in most cases, except for ART,
where it doubles the instructions retired per request for \textit{read-only} and \textit{read-heavy} workloads.

\subsubsection{Cycles per instruction (CPI)}
\label{results:overall:cpi}

\Cref{CPI} shows the cycles spent retiring an instruction on average across all the experiments.
The smaller this value is the better,
indicating that the processor core can exploit more instruction-level parallelism
issuing and retiring more instructions in a cycle (\Cref{background:uarch}). 
On the server we are using for these experiments,
the minimum for this value is 0.25 as mentioned in \Cref{setup:metrics}.

Data-intensive applications, especially transaction processing applications,
are famous for exhibiting low instruction-level parallelism (\Cref{background:related}).
The reason for this, especially for modern in-memory-optimized systems,
is that frequent random data accesses lead to long-latency stalls trying to fetch data all the way down from main memory,
when they do not hit in the L1 data cache.
In addition,
the complex instruction footprint of traditional disk-based data management systems cause high front-end stalls.

ALEX, ART, and B+Tree are all in-memory index structures targeting transaction processing workloads.
They do not encapsulate all the instruction complexity for a data management system,
since they just represent part of a data management system.
On the other hand, index operations are the most common operations for most transactional workloads.
Therefore, our expectation is to observe similar micro-architectural trends
to in-memory transaction processing systems with heavy instruction footprint optimizations 
(as in HyPer results from \cite{SirinTPA16})
here and in the following subsections that detail the micro-architectural behavior.

Results from \Cref{CPI} demonstrate that ALEX exhibit the highest level of
instruction-level parallelism across all the experiments compared to ART and B+Tree.
This corroborates the fundamental design principle for learned tree indexes,
where a trade-off for increased computation is made instead of memory accesses \cite{bailisOpinion}.
Especially, with the \textit{insert-only} workload inserting consecutive keys (\Cref{CPI_160M} and \Cref{CPI_1.6B}),
ALEX comes very close to the theoretical minimum with a CPI value of 0.36.
This is reasonable as the large instruction footprint reported in \Cref{results:overall:instr}
is mainly computation heavy, which does not lead to too many data fetches from main-memory.
In addition, both model training and possible later append operations do not exhibit complex program logic
that could leave branch predictors and next-line prefetching in modern processors ineffective.

\begin{figure*}
\centering
\begin{subfigure}{0.3\linewidth}
\centering
\includegraphics[width=\linewidth]{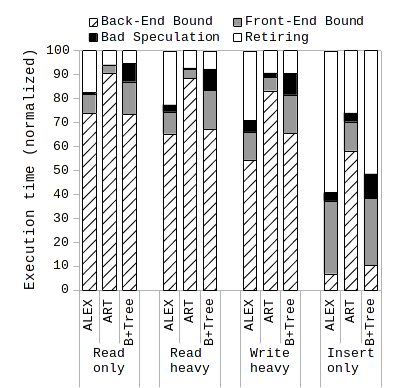}
\caption{Consecutive - 160 million keys}
\label{ExecutionCycles_160M}
\end{subfigure}%
\hspace{1em} 
\begin{subfigure}{0.3\linewidth}
\centering
\includegraphics[width=1\linewidth]{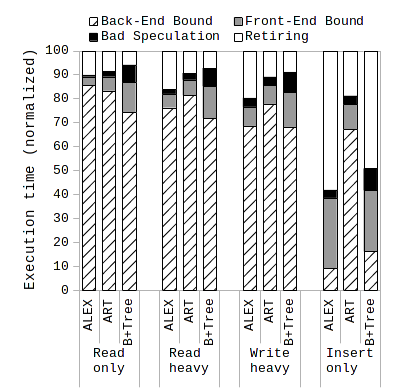}
\caption{Consecutive - 1.6 billion keys}
\label{ExecutionCycles_1.6B}
\end{subfigure}
\hspace{1em} 
\begin{subfigure}{0.3\linewidth}
\centering
\includegraphics[width=\linewidth]{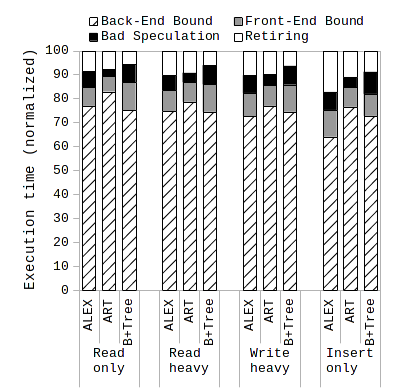}
\caption{Random - 1.6 billion keys}
\label{ExecutionCyclesR_1.6B}
\end{subfigure}
\caption{Breakdown of execution cycles (normalized).}
\label{ExecutionCycles}
\end{figure*}

\begin{figure*}
\centering
\begin{subfigure}{0.3\linewidth}
\centering
\includegraphics[width=\linewidth]{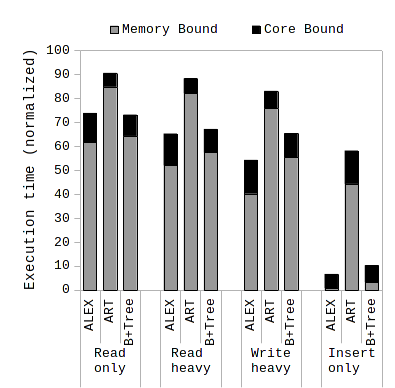}
\caption{Consecutive - 160 million keys}
\label{BackendStall_160M}
\end{subfigure}%
\hspace{1em} 
\begin{subfigure}{0.3\linewidth}
\centering
\includegraphics[width=1\linewidth]{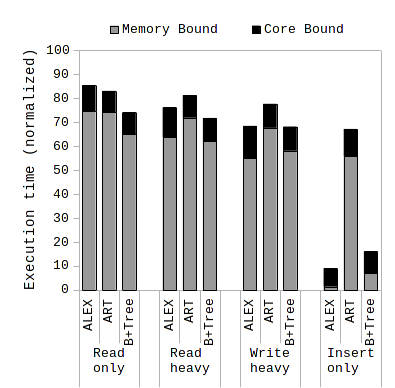}
\caption{Consecutive - 1.6 billion keys}
\label{BackendStall_1.6B}
\end{subfigure}
\hspace{1em} 
\begin{subfigure}{0.3\linewidth}
\centering
\includegraphics[width=\linewidth]{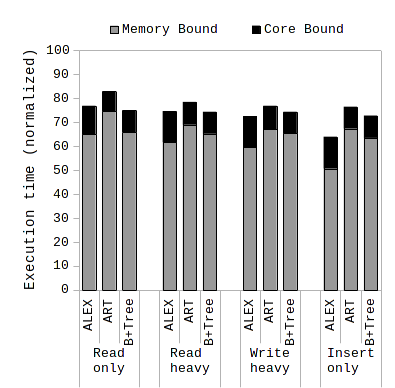}
\caption{Random - 1.6 billion keys}
\label{BackendStallR_1.6B}
\end{subfigure}
\caption{Breakdown of back-end stalls (normalized).}
\label{BackendStall}
\end{figure*}

For workloads with high percentage of reads or for random insert patterns,
all three indexes spend more cycles per instruction as a result of increased random data fetches in a unit of time.
This is a natural side-effect of a tree-index traversal,
which the results from the following subsections will further corroborate.

The reduction in CPI for \textit{read-only} and \textit{read-heavy} workloads
as the data size is increased (comparing \Cref{Instructions_160M} and \Cref{Instructions_1.6B})
also underlies the effect of traditional index traversal in ART and B+Tree.
The way these tree-based indexes are designed,
the level of the trees do not increase proportionally to the data size
avoiding proportional increase in random data access requests to main-memory.
Furthermore,
the traversal code is also a tight-loop with relatively small instruction footprint
that can be kept in the L1 instruction cache.
These two fundamental characteristics of tree-based structures
benefit the scalability with respect to data size for all three indexes.
On the other hand,
CPI for ALEX increases for the same scenario,
while still being smaller than the other two indexes as a result of the trade-offs mentioned above.
This is likely the impact of traversing a deeper hierarchy of models with the larger dataset.

We also see the positive impact of data locality for ART and B+Tree in the \textit{insert-only} workload
when comparing the \textit{consecutive} (\Cref{CPI_1.6B}) and \textit{random} (\Cref{CPIR_1.6B}) inserts.

Finally, overall, the trends in \Cref{ExecutionTime} and \Cref{CPI} match for the cases of the large dataset sizes,
and can be explained with similar reasoning.
On the other hand, for the small dataset size,
despite exhibiting a high CPI value, ART has more efficient execution time per request compared to the B+Tree,
which may come as a result of its lower instruction footprint per request in \Cref{Instructions}.

\subsection{Breakdown of execution cycles}
\label{results:bkExec}

\Cref{results:overall} demonstrated the top-level performance comparison across ALEX, ART, and B+Tree.
While interpreting the results from the top-level metrics,
we already hinted at certain causes at the micro-architectural-level
for the observed behavior.
Now, in this and the following two subsections,
we quantify such causes in more detail.

\Cref{ExecutionCycles} plots the breakdown of execution cycles into top-level micro-architectural components for all the experiments.
We normalize the cycles to 100\% and report the breakdowns in terms of percentages instead of absolute values,
since \Cref{results:overall:executiontime} already compared the execution time for all three indexes in absolute values.
Later, before closing our analysis of the experimental results,
\Cref{results:bkMB} also gives a detailed breakdown of the execution cycles over the absolute execution time
without normalizing, to circle back to the results presented earlier.

With the expected exception of the \textit{insert-only} workload with \textit{consecutive} pattern, 
\Cref{ExecutionCycles} emphasizes that majority of the execution time for all index structures goes to \textit{back-end} stalls. 
The time spent in \textit{retiring} instructions is relatively small,
as well as the stalls due to \textit{bad speculation} and \textit{front-end}.
This behavior is expected, since index operations are data access heavy,
leading to many random data accesses to main-memory 
(as \Cref{results:bkMB} will further underline). 

The portion of execution time ALEX spends in retiring instructions
increases as the workload exhibits more out-of-bound inserts
(\Cref{ExecutionCycles_160M} and \Cref{ExecutionCycles_1.6B})
as well as the \textit{front-end} component.
This is expected looking at the instructions required per request in \Cref{Instructions} for the same cases.

Overall,
ALEX spends a smaller percentage of its execution time in \textit{back-end} stalls compared to ART across all the experiments
and percentage of \textit{back-end} stalls is similar between ALEX and B+Tree.
However, B+Tree spends a higher portion of time in \textit{front-end} stalls and \textit{bad speculation}.
In terms of more stable or robust micro-architectural behavior, though,
ART's behavior does not change drastically across the experiments,
even with the extreme case of the \textit{insert-only} workload with \textit{consecutive} out-of-bound inserts.
ART is the most \textit{back-end bound} compared to the other two indexes.
This can be attributed to both its larger memory footprint (\Cref{results:overall:memoryfootprint}) and
small instruction footprint (\Cref{results:overall:instr}) increasing the pressure on the back-end of a processor.

Finally,
for the \textit{random} case illustrated in \Cref{ExecutionCyclesR_1.6B},
we do not see drastic changes in the behavior of the breakdown across different workload patterns.
Here the random data accesses are simply the core operation all the indexes perform, which leads to this outcome.

\subsection{Breakdown of back-end stalls}
\label{results:bkBE}

Since the execution cycles are mainly \textit{back-end bound} in \Cref{ExecutionCycles},
\Cref{BackendStall} breaks down the \textit{back-end} stalls further into
\textit{memory bound} and \textit{core bound} components for all the experiments.
\Cref{BackendStall} keeps the total execution time normalized at 100\%,
while the total of the \textit{memory bound} and \textit{core bound} components
sums up to the \textit{back-end bound} component from \Cref{ExecutionCycles}.

As expected, except for the case of out-of-bound inserts,
majority of the \textit{back-end} stalls are \textit{memory bound},
because of the random data accesses.
In addition, while small,
the \textit{core bound} component tends to be bigger for ALEX,
because of the more compute-heavy nature of learned indexes.
This may cause slightly increased pressure on the execution units in the back-end.

For the case of out-of-bound inserts (\Cref{BackendStall_160M} and \Cref{BackendStall_1.6B}),
ALEX spends almost no time in \textit{memory bound} stalls,
which is interesting to observe and
highlights the extremely compute-heavy nature of model re-training
triggered by this workload pattern.
B+Tree exhibits the impact of data locality in this case,
whereas ART's lower instruction and larger data footprint keeps the \textit{memory bound} component big.

Finally,
the trends across different dataset sizes and data insert patterns,
aligns with the insights we had for the CPI results in \Cref{CPI}, 
and can be attributed to the index traversal logic as explained in \Cref{results:overall:cpi}. 
To re-iterate,
the \textit{random} inserts (\Cref{BackendStallR_1.6B})
as well as \textit{random} read requests (\textit{read-only} and \textit{read-heavy}),
are bound by the impact of the random data accesses. 
Increasing the data size leads to more cache locality in both
instruction and data accesses for the tree-based index structures
that are not based on models such as ART and B+Tree.
On the other hand,
the increased number of models in ALEX with the increased data size leads to an increased \textit{memory bound} behavior.

\begin{figure*}
\centering
\begin{subfigure}{0.3\linewidth}
\centering
\includegraphics[width=\linewidth]{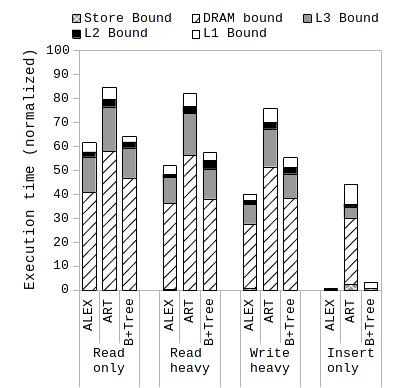}
\caption{Consecutive - 160 million keys}
\label{MemoryBoundStall_160M}
\end{subfigure}%
\hspace{1em} 
\begin{subfigure}{0.3\linewidth}
\centering
\includegraphics[width=1\linewidth]{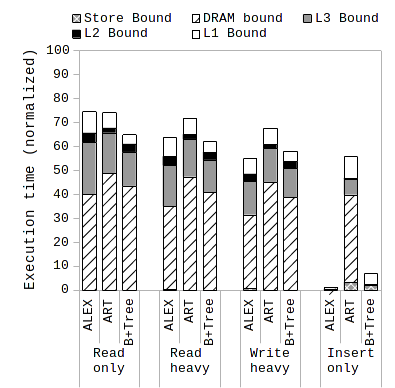}
\caption{Consecutive - 1.6 billion keys}
\label{MemoryBoundStall_1.6B}
\end{subfigure}
\hspace{1em} 
\begin{subfigure}{0.3\linewidth}
\centering
\includegraphics[width=\linewidth]{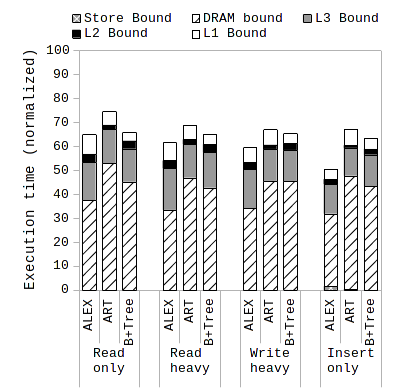}
\caption{Random - 1.6 billion keys}
\label{MemoryBoundStallR_1.6B}
\end{subfigure}
\caption{Breakdown of memory stalls (normalized).}
\label{MemoryBoundStall}
\end{figure*}

\begin{figure*}
\centering
\begin{subfigure}{0.3\linewidth}
\centering
\includegraphics[width=\linewidth]{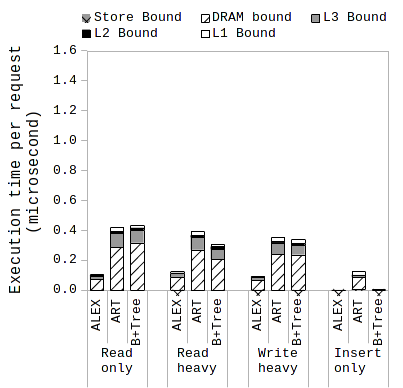}
\caption{Consecutive - 160 million keys}
\label{MemoryBoundStallTIME_160M}
\end{subfigure}%
\hspace{1em} 
\begin{subfigure}{0.3\linewidth}
\centering
\includegraphics[width=1\linewidth]{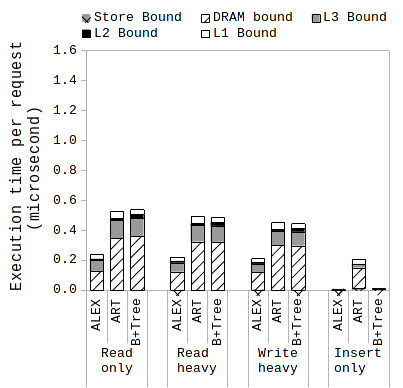}
\caption{Consecutive - 1.6 billion keys}
\label{MemoryBoundStallTIME_1.6B}
\end{subfigure}
\hspace{1em} 
\begin{subfigure}{0.3\linewidth}
\centering
\includegraphics[width=\linewidth]{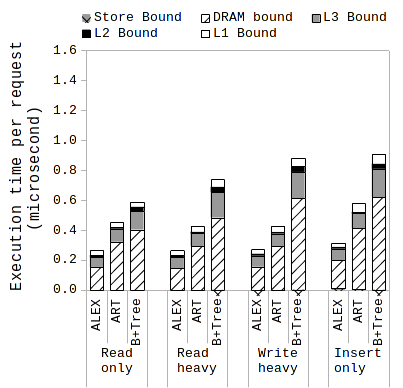}
\caption{Random - 1.6 billion keys}
\label{MemoryBoundStallTIMER_1.6B}
\end{subfigure}
\caption{Breakdown of memory stalls per request in $\mu$sec.}
\label{MemoryBoundStallTIME}
\end{figure*}

\subsection{Breakdown of memory stalls}
\label{results:bkMB}

Since the \textit{back-end bound} cycles are mainly \textit{memory bound} in \Cref{BackendStall},
in this section, we break the \textit{memory bound} stalls further into
where they come from within the memory hierarchy.
This section illustrates the results in two forms.
First, \Cref{MemoryBoundStall} plots this breakdown with respect to normalized execution time.
This means that the total of the presented components in \Cref{MemoryBoundStall}
sums up to the \textit{memory bound} component from \Cref{BackendStall}.
Then,
\Cref{MemoryBoundStallTIME} shows this breakdown with respect to the absolute values for the execution time from \Cref{ExecutionTime}
presenting the absolute time for various \textit{memory bound} stalls per request.
This also allows us to circle back to the top-level performance metrics we presented earlier this section.
The y-axes in \Cref{MemoryBoundStallTIME} is, therefore, kept the same as the results in \Cref{ExecutionTime} for better comparison.

\Cref{MemoryBoundStall} demonstrates that majority of the \textit{memory bound} stalls
are due to long-latency data misses from the last-level cache (L3),
which is labeled as \textit{DRAM bound}, for all indexes. 
This is followed by the data misses from L2 cache that hit in L3 cache,
which the \textit{L3 Bound} component represents.
In contrast, \textit{L2 Bound} component,
which shows stall time due to L1 data cache misses, 
is quite small as well as the \textit{L1 Bound} component,
which is mostly due to DTLB misses in our case.

This behavior of the memory stalls can be explained through both the hardware and software behavior.
On the one hand,
modern OoO processors are designed to overlap the latency for the L1 data misses for the most part, which is 7-8 cycles.
However, the latency for L2 and L3 misses are larger (17 and over 200 cycles, respectively).
Therefore, it becomes hard to overlap all the associated latency due to frequent random data accesses to L3 and DRAM.
On the other hand, the tree-based index traversal logic triggered by
the get/put/update requests to indexes in our (and typical transactional) workloads,
exhibit very low temporal and spatial locality
leading to many cache misses from all levels of the cache hierarchy.
Therefore, except for the accesses to the higher-levels of the index (e.g., root), 
most data accesses end up in DRAM. 

The case that shows different behavior is again the out-of-bound inserts in \Cref{MemoryBoundStall_160M} and \Cref{MemoryBoundStall_1.6B}.
ALEX and B+Tree exhibit almost no stalls due to memory accesses here as \Cref{results:bkBE} also covered.

Finally,
\Cref{MemoryBoundStallTIME} shows that, in absolute values,
ALEX spends the least amount of time in \textit{memory bound} stalls compared to ART and B+Tree,
which aligns with its normalized values as well.
On the other hand, ART and B+Tree spend similar amount of time in \textit{memory bound} stalls in \textit{consecutive} case,
while ART spends shorter time in \textit{memory bound} stalls in \textit{random} case,
despite the normalized values in \Cref{MemoryBoundStall}.
This is a result of ART's more efficient utilization of the \textit{front-end} component,
which is very small for ART in the normalized case, 
increasing the impact and portion of the \textit{memory bound}.
Therefore,
it is important to have both the normalized and absolute perspectives
while breaking down the execution cycles into micro-architectural components.

\section{Conclusion}
\label{conclusion}

In this paper,
we performed a micro-architectural analysis of a learned index, ALEX, 
and compared its behavior to two tree-based indexes that are not based on learned models, ART and B+Tree. 
Our study complements existing experimental comparison results for learned indexes
by providing a lower-level view of how well learned indexes
utilize the front-end, back-end, memory hierarchy, etc. of modern commodity servers with OoO processors.

We used a variety of workloads with different \% of read and write requests
as well as different data insert and access patterns and sizes. 
In summary, our results show that,
long-latency data misses is at the core of ALEX
similar to the other two indexes.
This is expected as the short get/put requests 
cause frequent random data accesses that lack locality,
which stress the memory hierarchy of a processor.
On the other hand, 
ALEX exhibits higher instruction-level parallelism 
and fewer stalls cycles in general.
This is a result of the fact that
learned indexes trade-off increased computations to reduced memory accesses,
which could boost performance if the overall computation cycles 
are cheaper with respect to memory access latency \cite{bailisOpinion}.
In the case of modern OoO processors,
computation cycles are indeed cheaper 
if the instruction footprint is not complex,
i.e., does not have too many dependencies.
Even if ALEX's instruction footprint drastically increases 
in the case of inserting out-of-bound keys,
these instructions are less stall prone,
since they exhibit very high instruction-level parallelism.

Going forward,
investigating the behavior for the same index structures 
for workload patterns that dynamically change 
over time during a single run would be interesting
to observe the impact of various insert heavy periods 
on later read heavy periods of execution.
In addition,
in this study we mainly looked at uniformly random data access patterns.
Adopting a variety of access patterns as well as more complex insert behavior would also be an interesting future analysis.

\subsection*{Acknowledgements}
The authors would like to thank Intel DevCloud \cite{devcloud} for providing the server 
and the maintainers of the open-source codebases used used by the experiments in this paper.

\bibliographystyle{ACM-Reference-Format}
\bibliography{main}

\balance

\end{document}